\documentclass[a4paper]{article}
\usepackage[T1]{fontenc}
\usepackage{authblk}
\usepackage{graphicx}

\title{New developments in Micromegas Microbulk detectors}
\author[1]{F.J.~Iguaz}
\author[2]{S.~Andriamonje}
\author[1]{F.~Belloni}
\author[1]{E.~Berthoumieux}
\author[2]{M.~Calviani}
\author[3]{T.~Dafni}
\author[2]{R.~De Oliveira}
\author[1]{E.~Ferrer-Ribas}
\author[1]{J.~Gal\'an}
\author[3]{J.A.~Garc\'ia}
\author[1]{I.~Giomataris}
\author[2]{C.~Guerrero}
\author[1]{F.~Gunsing}
\author[3]{D.C.~Herrera}
\author[3]{I.G.~Irastorza}
\author[1]{T.~Papaevangelou}
\author[3]{A.~Rodr\'iguez}
\author[3]{A.~Tom\'as}

\affil[1]{IRFU, Centre d'\'Etudes Nucl\'eaires de Saclay (CEA-Saclay), Gif-sur-Yvette, France}
\affil[2]{European Organization for Nuclear Research (2), Gen\`eve, Switzerland}
\affil[3]{Laboratorio de F\'isica Nuclear y Astropart\'iculas,
Universidad de Zaragoza, Spain}

\begin{document}

\maketitle

\begin{abstract}
A new Micromegas manufacturing technique, based on kapton etching technology, has been recently developed, improving the uniformity and stability of this kind of readouts. Excellent energy resolutions have been obtained, reaching 11\% FWHM for the 5.9 keV photon peak of $^{55}$Fe source and 1.8\% FWHM for the 5.5 MeV alpha peak of the $^{241}$Am source. The new detector has other advantages like its flexible structure, low material and high radio-purity. The two actual approaches of this technique will be described and the features of these readouts in argon-isobutane mixtures will be presented. Moreover, the low material present in the amplification gap makes these detectors approximate the Rose and Korff model for the avalanche amplification, which will be discussed for the same type of mixtures. Finally, we will present several applications of the microbulk technique.
\end{abstract}

\section{The microbulk technology}
\label{sec:microbulk}
Micromegas (for MICRO MEsh GAseous Structure) is a parallel-plate detector invented by I. Giomataris in 1995 \cite{Giomataris:1995fq}. It consists of a thin metallic grid (commonly called mesh) and an anode plane, separated by insullated pillars. Both structures define a very little gap (between 20 and 300~$\mu$m), where primary electrons generated in the conversion volume are amplified, applying moderate voltages at the cathode and the mesh. This technology has good properties like its high granularity, good energy and time resolution, easy construction, little mass and gain stability.

\medskip
First readouts were built screwing two different frames: the anode plane and the metallic grid, where kapton pillars were electroformed. By applying voltages to both structures, the intense electric field pulled down the mesh and the flatness was thus defined by the height of the pillars, which had an accuracy better than 10 $\mu$m. The good flatness and parallelism between the anode and the mesh was obtained only if the delicate operation of screwing was successful. To avoid this operation, two different technologies (called bulk \cite{Giomataris:2006yg} and microbulk \cite{Adriamonje:2010sa}) have been developed so that the readout plane and the mesh formed a single integrated structure.

\begin{figure}[htb!]
\centering
\includegraphics[width=60mm]{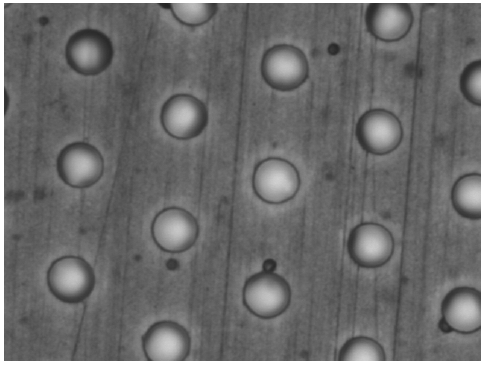}
\hspace{10mm}
\includegraphics[width=60mm]{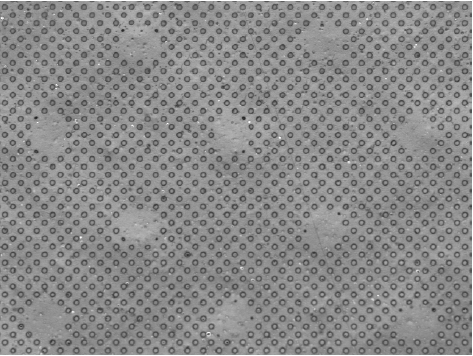}
\caption{\it Left: A view of the mesh of a conventional microbulk readout of 50$\mu$m gap. The holes have a diameter of 40$\mu$m, a pitch of 100$\mu$m and a triangular pattern. Right: A view of the mesh of a pillars microbulk of 50$\mu$m gap. Its holes have the same diameter and pitch as the former one but a square pattern. The pillars that define the gap have a diameter of 400$\mu$m and a pitch of 1 mm.}
\label{fig:mesh}
\end{figure}

\medskip 
In the microbulk technology, the raw material is a thin flexible polyimide foil with a thin copper layer on each side. The foil is glued on top of a rigid substrate that provides the support of the micro-structure and carries anode strips or pixels. In the first step, a thin photoresistive film is laminated on top of the kapton foil and is insolated by UV light to produce the required mask. The copper is then removed by a standard lithographic process, the non-insulated places producing a pattern of a thin mesh. One example is shown in figure \ref{fig:mesh} (left). The polyimide is then etched and partially removed in order to create tiny pillars in the shadow part below the copper mesh.

\medskip 
In a second approach of this technology, the polyimide is completely removed except where small pillars are created (about 100 $\mu$m in diameter) with a pitch of 1 mm. This process requires that an additional insolating spot (about 200 $\mu$m) is formed during the insolation process, leaving after the lithographic process a copper spot of 200 $\mu$m of diameter. By increasing the duration of the etching process, the polyimide under the mesh is removed completely except the space below these copper spots, as it is shown in figure \ref{fig:mesh} (right). This technique reduces the capacity and the total noise of the readout.

\medskip
For both techniques, the amplification gap is more homogeneous and the mesh geometry has a better quality than the classical Micromegas, reducing the avalanche fluctuations and improving the energy resolution, as later described. It also allows to decrease the border regions reducing the dead zone, which allows its application in imaging with x-rays or neutrons. Moreover, the readout is made of kapton and copper, two of the materials with the lowest levels of radiopurity \cite{Cebrian:2011sc}. On the other hand, microbulk readouts are less robust that the former ones and the maximum size is today 30 cm., limited by the actual equipment.

\section{Characterization in argon-isobutane mixtures}
\label{sec:ariso}
Two microbulk detectors with a gap of 50~$\mu$m have been characterized in argon-isobutane mixtures to study their general performance and the possible differences between both construction techniques. The first one was built with the conventional procedure, where only the kapton of the hole is removed. For the second one, all the kapton between the mesh and the anode plane was removed, except for few separated pillars that keep apart the two structures. Both readouts are circular, with a diameter of 35 mm, and have a single non-segmented anode covering all the area. The specific references of the detectors are the M50.70.35.04 and the M50.50.25.01, which respectively correspond to a pitch distance of 70 and 50~$\mu$m and a hole diameter of 35 and 25~$\mu$m. Althought their dimensions are different, the ratios between the pitch distance and the hole diameter are the same, which implies no difference in the electron transmission curve due to the specific readout design \cite{Giganon:2011ag}. These particular readouts showed an energy resolution of 11.6\% and 12.2\% FWHM at 5.9 keV in Ar~+~5\%~iC$_4$H$_{10}$.

\subsection{Setup description}
\begin{figure}[htb!]
\centering
\includegraphics[width=70mm]{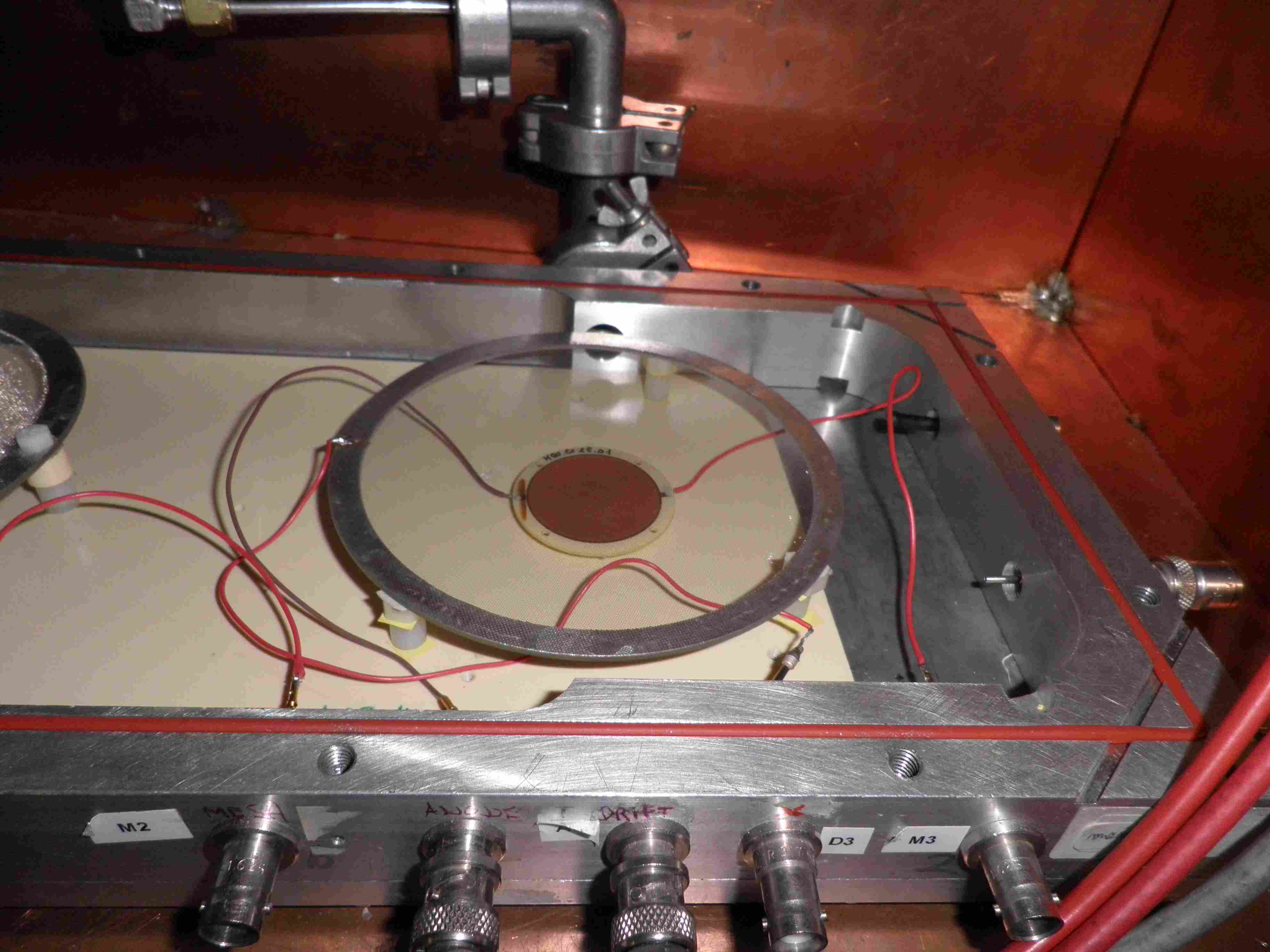}
\hspace{5mm}
\includegraphics[width=80mm]{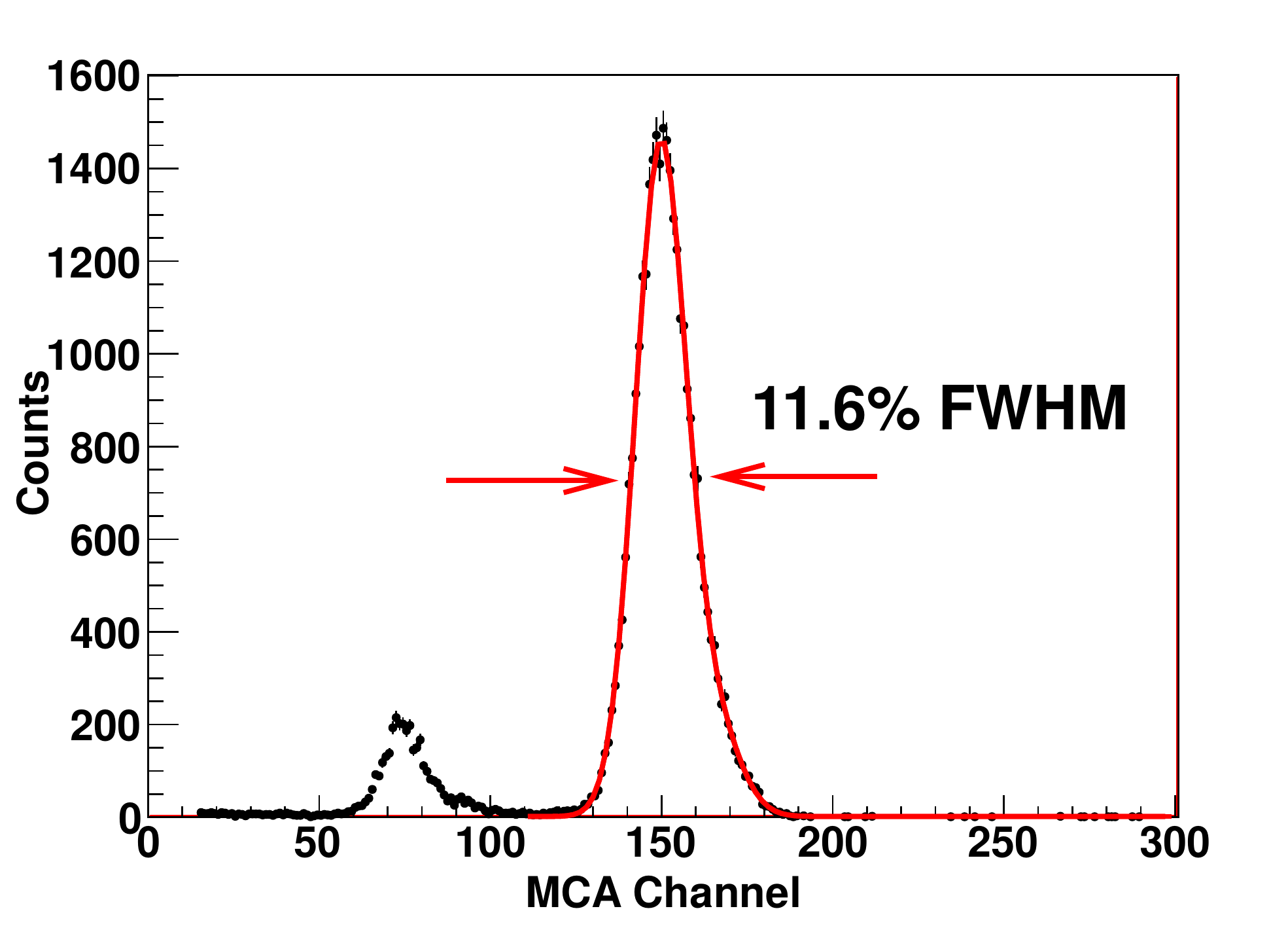}
\caption{\it Left: A view of one microbulk readout installed in the chamber, described in detailed in the text. Right: Iron energy spectrum obtained with the conventional readout in Ar+5\%iC$_4$H$_{10}$ at atmospheric pressure. The left peak corresponds to the escape peak of the argon.}
\label{fig:setup2}
\end{figure}

The vessel was specifically designed to characterize a maximum of three micromegas detectors (not necessary microbulk) in the same gas conditions, with a maximum surface of $50 \times 50$ mm$^2$. A general view of the setup is shown in figure \ref{fig:setup2} (left), where the top cap has been removed. The chamber consists in an aluminium box, with an internal volume of $60 \times 165 \times 30$ mm$^3$ and a wall thickness of 8 mm. The vessel is equipped with two gas entrances to circulate the desired gas. Readouts are placed on a plastic support situated inside the chamber and that electrically isolates the detectors from the inner walls. For each readout, there is a mesh frame screwed to the plastic plate, that is used as a drift cathode. The distance between the drift cathode and the detector is 10~mm. The drift and micromegas voltages are respectivelly extracted by a high voltage (SHV) and two signal (BNC) feed-throughs for each readout. On the top cap of the vessel, several holes have been made to calibrate the readouts by gamma sources situated outside. This holes are covered by an aluminized mylar film, that is scotched to the inner wall of the cap with kapton and creates the gas-tightness of the chamber.

\medskip
During at least 1 hour, a flow of the desired argon-isobutane mixture circulated by the vessel. The readout was then tested with an iron $^{55}$Fe source (x-rays of 5.9 keV) keeping the same gas flow. The mesh voltage was typically varied from 200 to 500~V and the drift one from 280 to 5000~V, reaching higher values for higher concentrations of isobutane. Both voltages were powered independently by a CAEN N471A module. A negative signal was induced in the mesh and read out by an ORTEC 142C preamplifier. The preamplifier output was fed into an ORTEC 472A spectroscopy amplifier and subsequently into a multichannel analyzer AMPTEK MCA-8000A for spectra building.

\subsection{Results}
\begin{figure}[htb!]
\centering
\includegraphics[width=80mm]{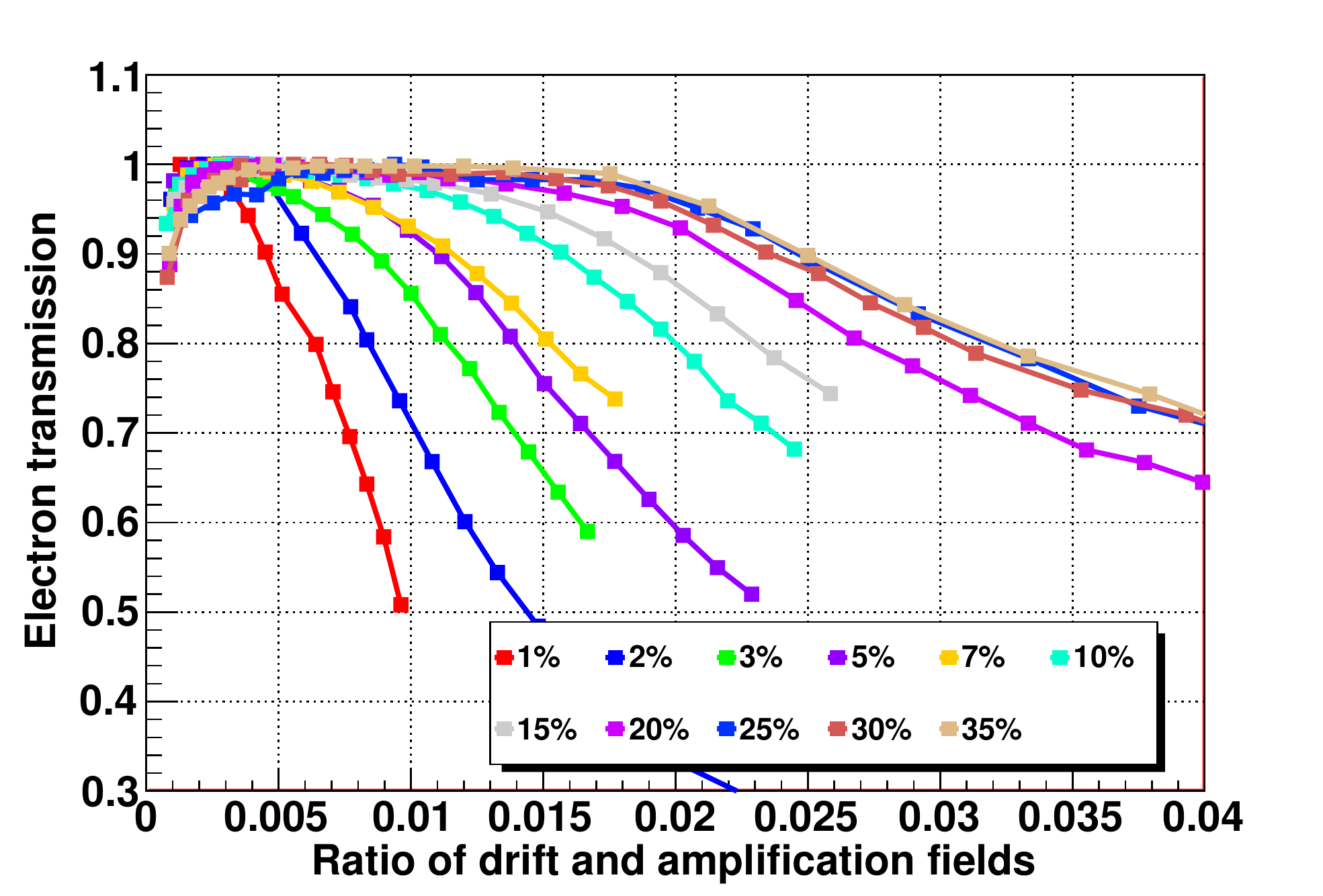}
\includegraphics[width=80mm]{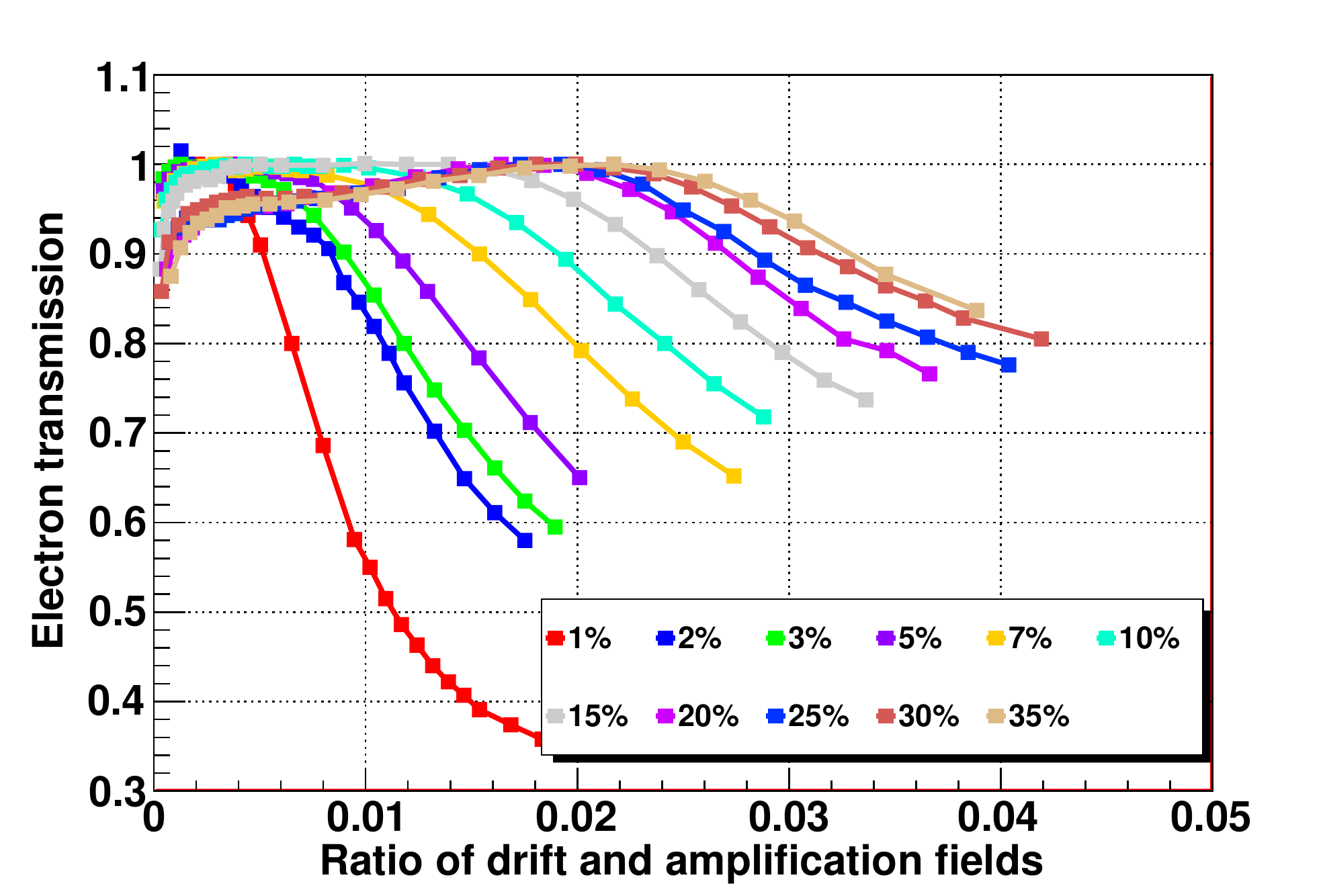}
\caption{\it Dependence of the electron transmission with the ratio of drift and amplification fields for a conventional (left) and pillars readout (right) in argon-isobutane mixtures between 1 and 35\%. The peak positions have been normalized with the maximum of each series. The percentage of each series correspons to the isobutane concentration.}
\label{fig:TransArIso}
\end{figure}

The drift voltage was firstly varied for a fixed mesh voltage to obtain the electron transmission curve, shown in figure \ref{fig:TransArIso} for the conventional (left) and the pillars readout (right). Microbulk detectors show a plateau of maximum electron transmission for a range of ratios of drift and amplification fields. For higher drift fields, the mesh stops being transparent for the primary electrons generated in the conversion volume and both the gain and the energy resolution degrades. As shown for the conventional readout, the specific maximum value for the ratio depends on the percentage of isobutane, being wider the plateau for higher quantities of the quencher. This fact is in agreement with the shorter plateau observed in pure gases \cite{Iguaz:2010fj} and seems to be correlated with diffusion coefficients \cite{Chefdeville:2009mc}. In contrast, the electron transmission curve of the pillars readout has a different behaviour for isobutane percentages greater than 20~\%. There is no clear plateau but a steady increase of about 5~\% between ratios of 0.005 and 0.02.

\begin{figure}[htb!]
\centering
\includegraphics[width=80mm]{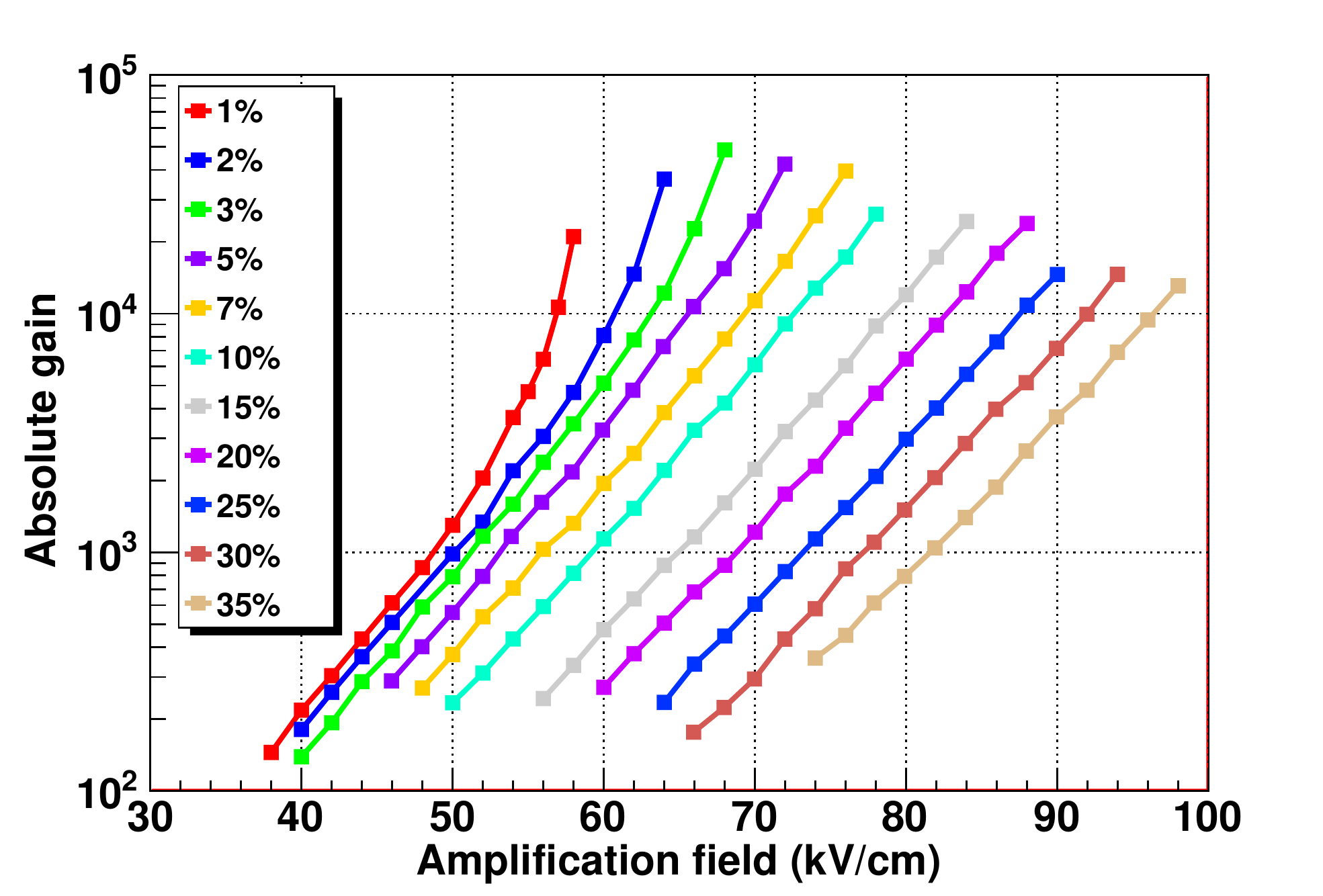}
\includegraphics[width=80mm]{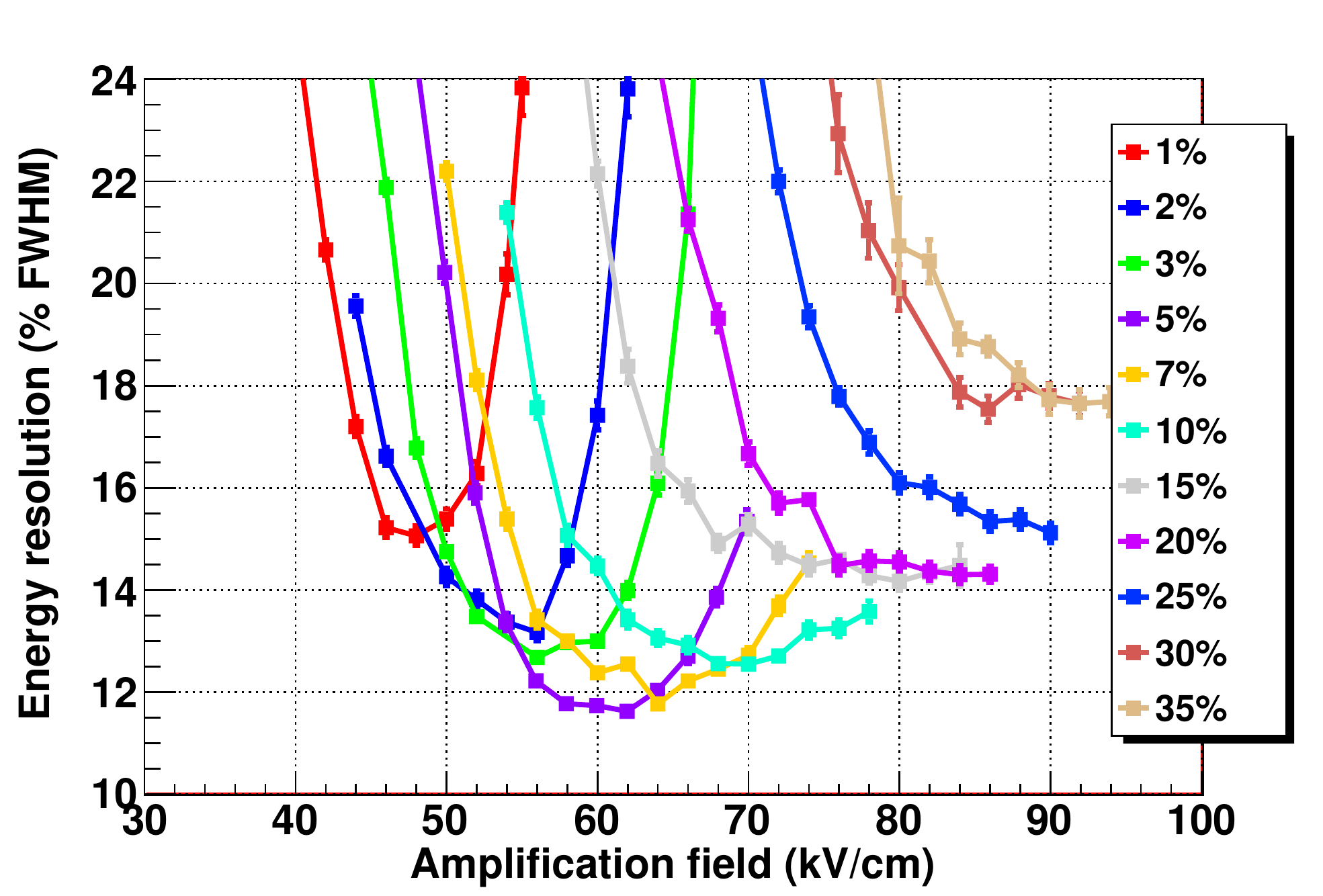}
\caption{\it Left: Dependence of the absolute gain with the amplification field for a conventional readout in argon-isobutane mixtures between 1 and 35\%. The maximum gain of each curve was obtained just before the spark limit. Right: Dependence of the energy resolution with the amplification field for a conventional readout in argon-isobutane mixtures between 1 and 35\%.}
\label{fig:GainClassic}
\end{figure}

\medskip
After having studied the mesh transparency, the ratio of drift and amplification fields was chosen so as the mesh showed the maximum electron transmission. In the case of the pillars readout, the nearest point to the maximum was taken, keeping a good energy resolution. The dependence of the peak position with the mesh voltage generates the gain curves, shown in figure \ref{fig:GainClassic} (left) for the conventional readout. Both detectors reach gains greater than $2 \times 10^4$ before the spark limit for all mixtures. Apart from that, the gain curves for low quantities of isobutane show a deviation from the Rose and Korff gain model at high amplification fields. This over-exponential behaviour is due to the low quencher concentrations, which cannot avoid the production of secondary avalanches \cite{Bronic:1998ikb}.

\medskip
The dependence of the energy resolution with the amplification field for the conventional readout is shown in figure \ref{fig:GainClassic}. The same curves were obtained for the pillars readout. For each mixture, there is a range of fields were the energy resolution is constant. This range is wider for mixtures rich in isobutane. For low fields (and gain), this parameter degrades because the signal starts being comparable with noise. For high fields, the resolution worsens due to the increase of the gain fluctuations by th UV photons generated in the avalanche. This effect disappears at higher quencher concentrations because these photons are better absorbed. However, the energy resolution worsens for higher isobutane proportions due to a reduction of the mean energy of the avalanche electrons, which increases the number of scattering processes over the ionization ones and the gain fluctuations \cite{Schindler:2010hs}.

\section{New gas mixtures for sub-keV applications}
\label{sec:neiso}
Micromegas detectors have been tipically operated in argon-isobutane mixtures between 2 and 5\% for a wide range of applications. This gas is well adapted for measurements in the 1-10 keV range, providing very good energy resolution and gains up to $2 \times 10^4$. Other gas mixtures are being studied to increase micromegas sensitivity in the sub-keV region, which could allow its application in synchroton radiation and dark matter searches where the low energy threshold is crucial. For lowering the energy threshold, the signal to noise ratio must be increased and higher gains are needed. One option considered is the replacement of argon by neon because the total charge per single avalanche is increased and it approaches the Rather limit ($\approx 10^8$ electrons).

\begin{figure}[htb!]
\centering
\includegraphics[width=83mm]{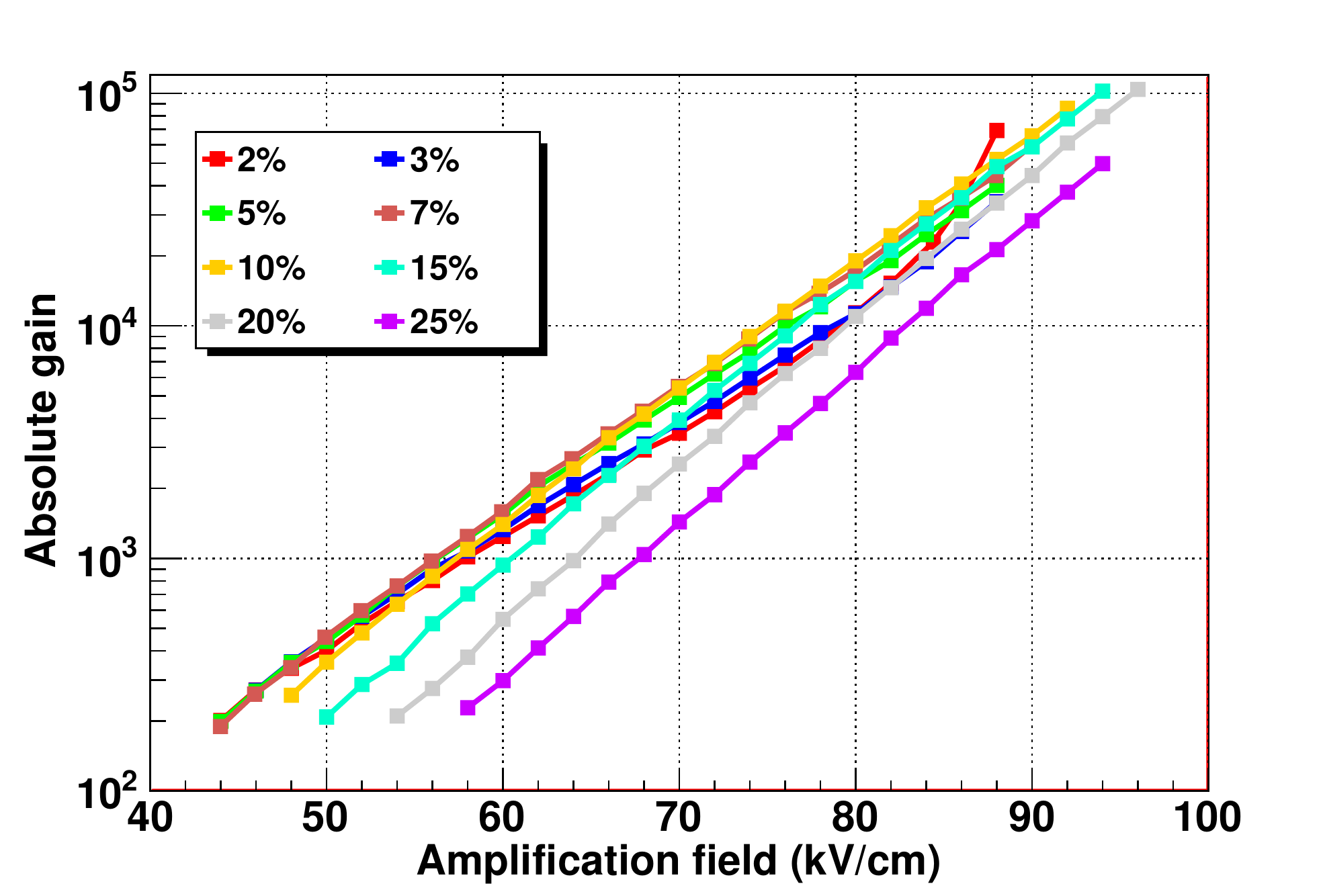}
\includegraphics[width=80mm]{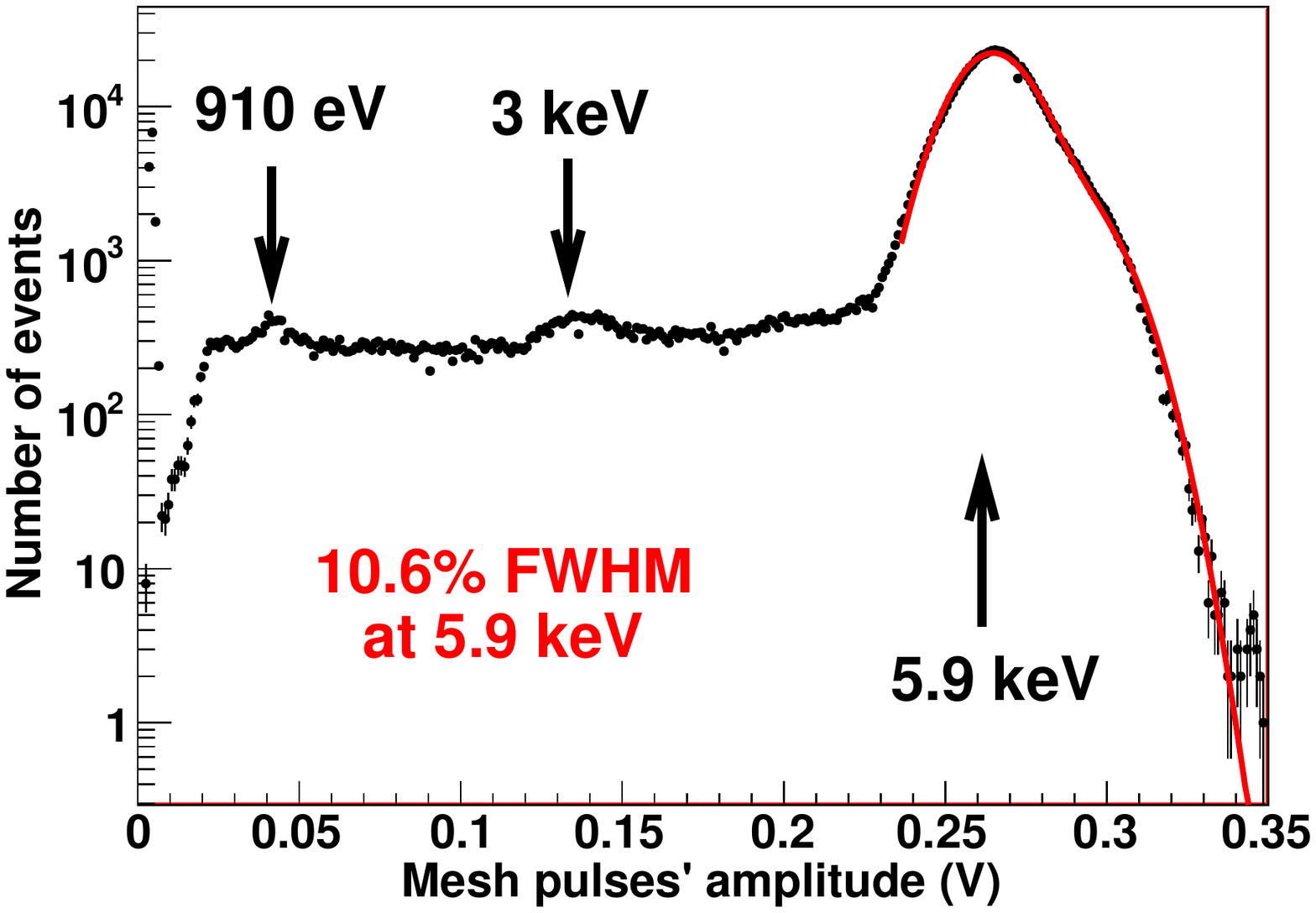}
\caption{\it Left: Dependence of the absolute gain with the amplification field for a conventional readout in neon-isobutane. The maximum gain of each curve was obtained just before the spark limit. Right: Energy spectrum generated by the mesh pulses of a conventional readout irradiated by a $^{55}$Fe source in Ne+7\%iC$_4$H$_{10}$ and a gain of $2 \times 10^4$. Note than y-axis is in logarithmic scale. The energy resolution at 5.9 keV is 10.6\% FWHM.}
\label{fig:GainNeIso}
\end{figure}

\medskip
The same setup and procedure described in former section was used to charaterize the conventional readout in neon-isobutane mixtures from 2 to 25\%. As in argon-based mixtures, the electron transmission curves showed a plateau, which widened for higher quencher concentrations. As shown in figure \ref{fig:GainNeIso} (left), gains over $5 \times 10^4$ were reached for all mixtures, reaching values around $10^5$ for isobutane proportions between 10 and 20\%. For different mixtures and voltages, mesh pulses were acquired by a LeCroy WR6050 oscilloscope. In an offline analysis, the pulses features were calculated and the energy spectrum was generated by the pulses' amplitude, as the one shown in figure \ref{fig:GainNeIso} (right). Apart from the main iron peaks at 5.9 and 6.4 keV, two others peaks with lower energy appears at 910 eV (the neon escape peak) and 3 keV (generated by the isobutane). The energy threshold is situtated at 400 eV, which points out that lower energies could be detected.

\begin{figure}[htb!]
\centering
\includegraphics[width=80mm]{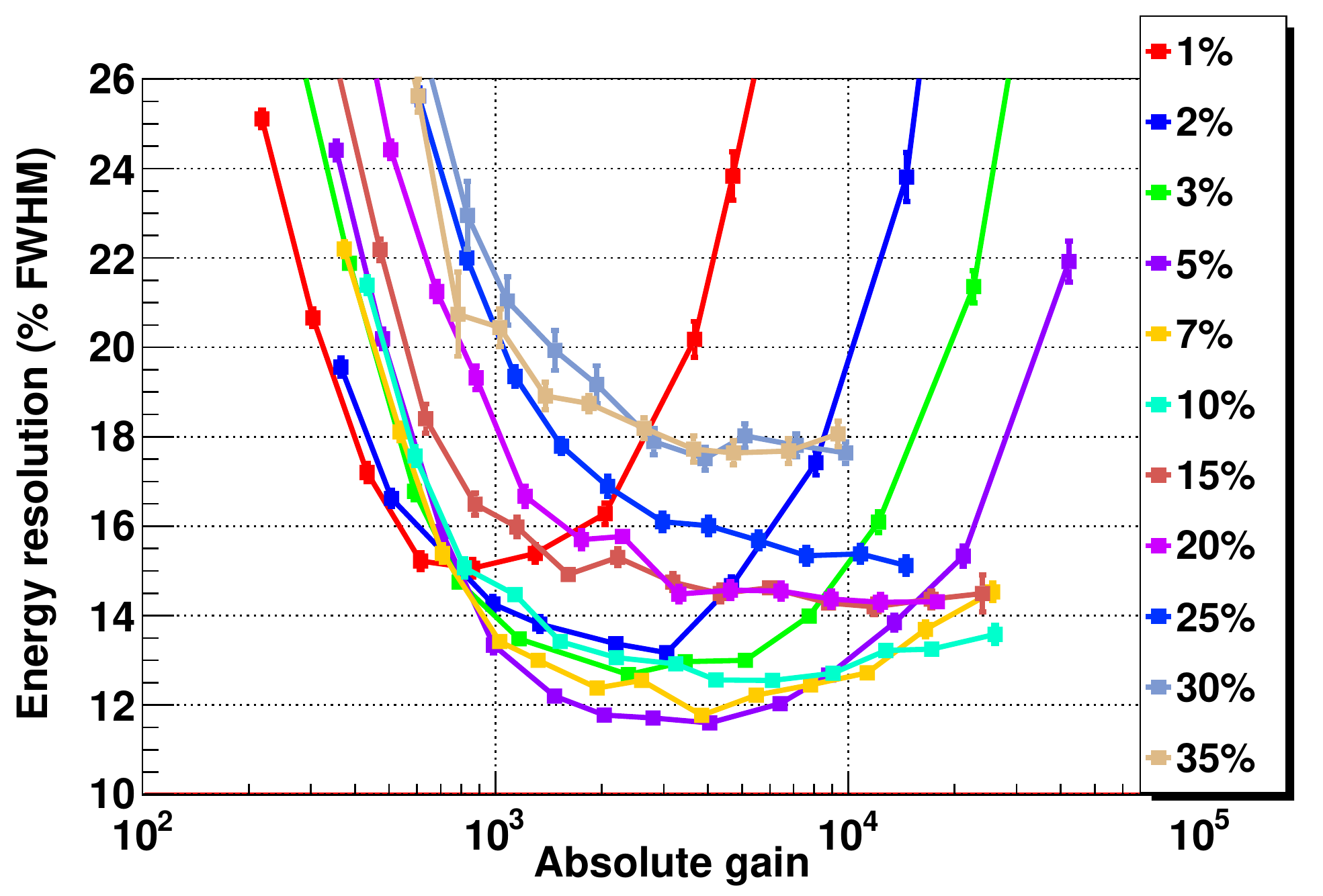}
\includegraphics[width=80mm]{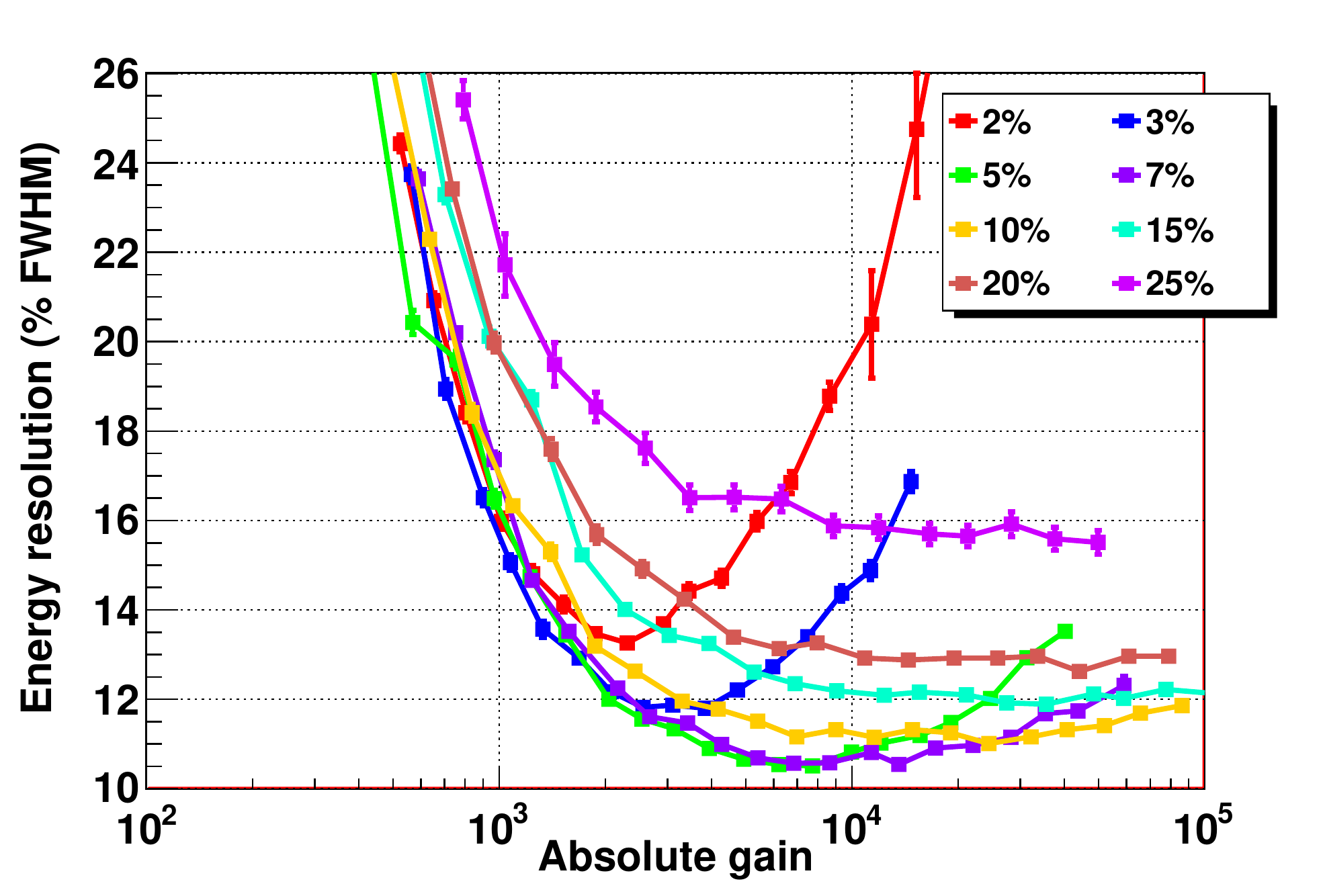}
\caption{\it Dependence of the energy resolution with the absolute gain for a conventional readout in argon- (left) and neon-isobutane mixtures (right). The maximum gain of each curve was obtained just before the spark limit.}
\label{fig:NeIso}
\end{figure}

\medskip
An improvement of the energy resolution was also observed for this mixtures, passing from 11.6\% FWHM for Ar+5\%iC$_4$H$_{10}$ down to 10.5\% FWHM for Ne+7\%iC$_4$H$_{10}$. This fact cannot be explained by the primary ionization but by the detector's avalanche. In neon-based mixtures, less primary electrons are generated due to its higher mean electron-ion pair energy (36.4~eV for neon and 26.3~eV for argon \cite{Blum:1994wb}), which would worsen the energy resolution, even if the Fano factor is slightly better (0.17 for neon, 0.22 for argon). However, less avalanche fluctuations are presented in light gases like helium or neon due to their higher ionization yield \cite{Schindler:2010hs}. The energy resolution is also better at higher gains, as shown in figure \ref{fig:NeIso}, where the dependence of the energy resolution with the gain of both mixtures is compared. At a gain of $5 \times 10^4$, an energy resolution of 11.4\% FWHM can be obtained in Ne+10\%iC$_4$H$_{10}$.

\medskip
Another research line is the replacement of the isobutane by another quencher. The conventional readout has already been tested in argon-cyclohexane mixtures, reaching gains up to $5 \times 10^4$. However, values for energy resolution are similar than in isobutane mixtures and they worsen fastly for higher gains. Other envisaged quencher is ethane, which could produce an increase of the primary ionization as its ionization potential (11.65 eV) is the nearest to the first argon metastables levels \cite{Agrawal:1988pca}. This effect is commonly known as Penning mixture.

\section{Applications}
\label{sec:app}
\subsection{CAST: a solar axion experiment}
The 2 Axion Solar Telescope (CAST) \cite{KZioutas:2005kz, Andriamonje:2010sa2, EArik09} uses a prototype of a superconducting LHC dipole magnet to convert axions into detectable X-ray photons. Axions are pseudoscalar particles that appear in the Peccei-Quinn solution of the strong CP problem and are candidates to Dark Matter. The Sun could produce a big flux of axions via the Primakoff conversion of plasma photons. These particles could then couple to a virtual photon provided by the magnetic transverse field of the CAST magnet, being transformed into a real photon that carries the energy and momentum of the original axion. The magnet follows the Sun twice a day during the sunset and sunrise for a total time of 3 hours per day.

\begin{figure}[htb!]
\centering
\includegraphics[width=60mm]{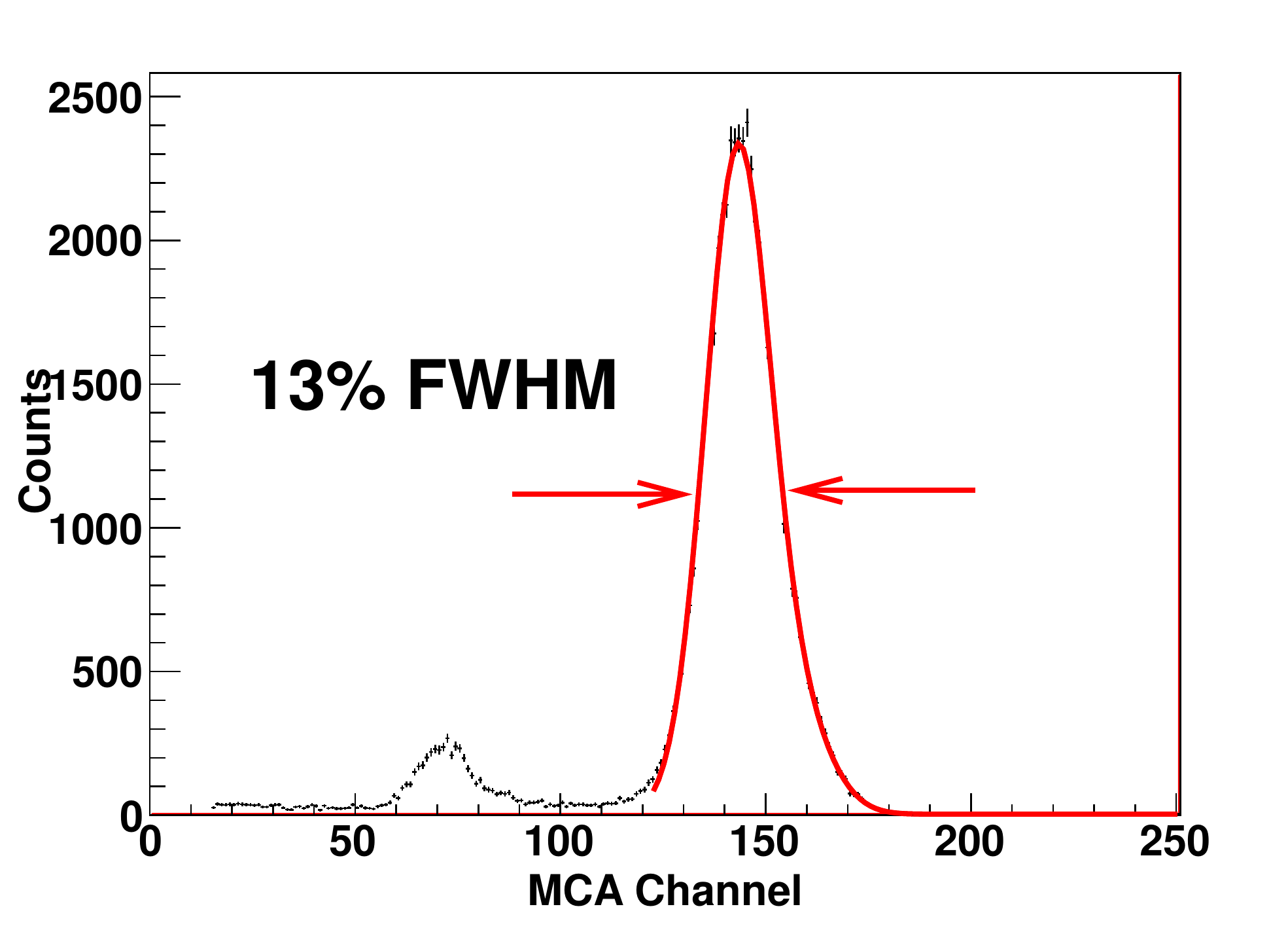}
\hspace{5mm}
\includegraphics[width=30mm]{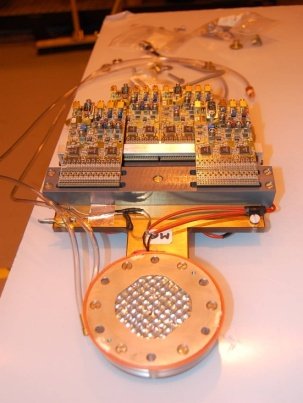}
\hspace{5mm}
\includegraphics[width=60mm]{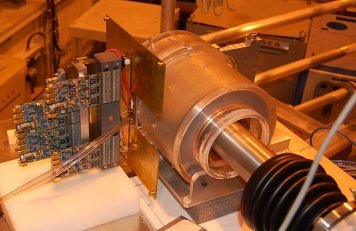}
\caption{\it Left: Energy spectrum obtained illuminating all the active area of the CAST-M16 detector in Ar+2.3\% isobutane at atmospheric pressure. Center: A photo of the CAST detector. The active area is situated at the front part and is covered with a stainless steel window and a plexiglass piece. The strips are read by four Gassiplex cards situated at the rear part. Right: The circular lead shielding that surrounds the readout and the stainless steel tube that comes out from the magnet bore.}
\label{fig:CAST}
\end{figure}

\medskip
Four X-ray detectors are installed at the ends of the 10 m long magnet in order to search for photons from Primakoff conversion. Since 2008, three microbulk Micromegas detectors are being used \cite{SAune:2009sa}, replacing a conventional Micromegas and a TPC. The readouts' anode consists of square pads interconnected in the diagonal directions through vias in two extra back-layers, producing a strip pitch of 550 $\mu$m. The energy resolution of these readouts reaches values as good as 13\% FWHM at 5.9 keV in argon-isobutane mixtures, as shown in figure \ref{fig:CAST} (left). The readout is situated in a chamber, shown in figure \ref{fig:CAST} (center), formed by a cylindrical plexiglas body and a stainless steel strong back that works as drift plate and is also used for the coupling with the magnet bore. The chamber is shielded by two inner layers of copper and archeological lead, as shown in figure \ref{fig:CAST} (right). More details are given in \cite{Tomas:2011at, Iguaz:2011fi}.

\subsection{Double beta decay experiments}
The measurements of the effective neutrino mass is one of the main goals of Neutrino Physics. The observation of the neutrinoless double beta decay ($0\nu\beta\beta$) would provide a good estimation of the mass and would confirm the Majorana nature of neutrinos. The actual setups being built will have a detector mass around 100 kg and will be sensitive to neutrino masses down to 50-100 meV. For this objective, the background level must below $10^{-3}$ c/keV/kg/yr at the $Q_{\beta\beta}$ and the energy resolution must be good enough to separate the tail of the $2\nu\beta\beta$ distribution from the neutrinoless peak. These two conditions are fullfilled by the microbulk technology \cite{Cebrian:2010sc2} and have motivated its feasibility study in a high pressure Xenon TPC \cite{Dafni:2010td, Iguaz:2010fi}, under the framework of the NEXT collaboration \cite{Granena:2011fg}.

\begin{figure}[htb!]
\centering
\includegraphics[width=80mm]{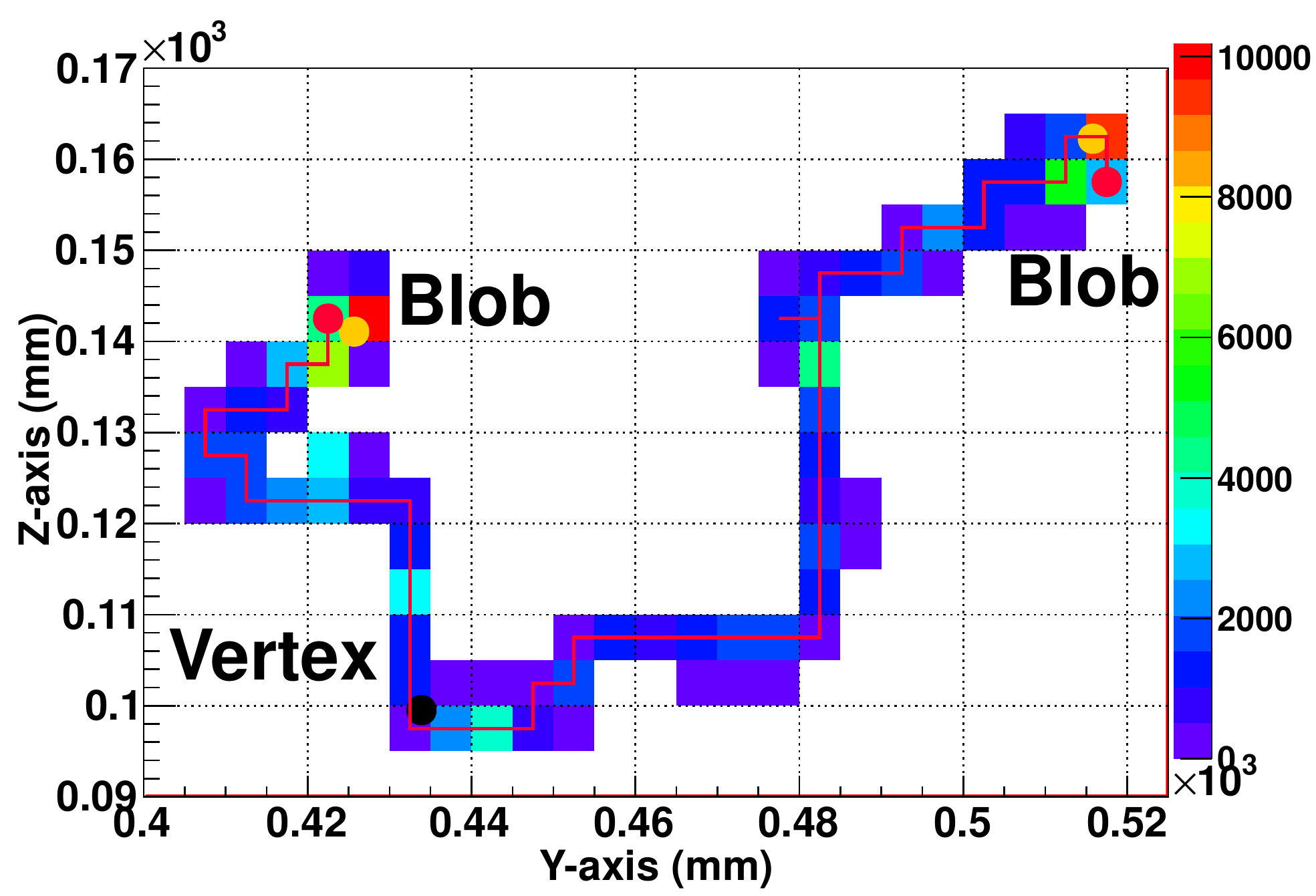}
\includegraphics[width=80mm]{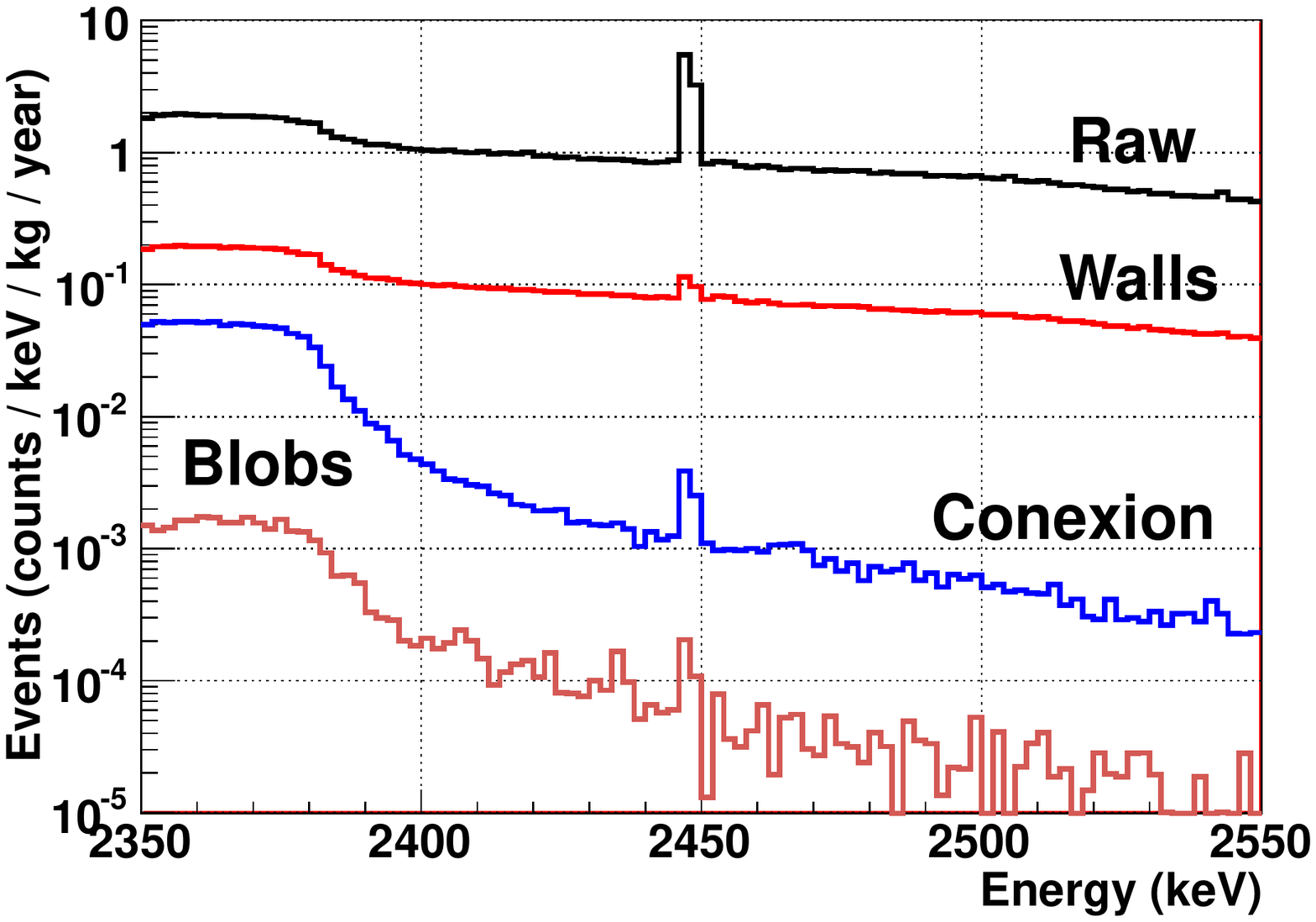}
\caption{\it Left: A $^{136}$Xe neutrinoless double beta decay event as it could be detected in a high pressure Xenon TPC at 10 bar equipped with a pixelized microbulk readout. The event consists in two electrons of 1.11 and 1.35 MeV emitted from the black point named vertex. Both electrons have a big charge deposition at the end of their tracks, generally called blob. Right: Estimated background level around 2458 keV produced by a 2 cm-thick copper veseel in a 100 kg high pressure Xenon TPC at 10 bar after having applied sucessive selection criteria: none (black line), fiducial volume (red line), only one connection (blue line) and two blobs at both ends (brown line).}
\label{fig:DB}
\end{figure}

\medskip
Microbulk detectors show an excellent energy resolution in pure xenon, down to 3\% FWHM at 2458 keV (the $Q_{\beta\beta}$ of $^{136}$Xe) at 10 bar and gains greater than $10^2$ \cite{Balan:2011cb}. Moreover, their high granularity allows a detailed image of the event topology, which can be used to discriminate signals from background events with high efficiency. The expect signal consists in the emission from a common vertex of two electrons that share the $Q_{\beta\beta}$. One example, as it would be registered by a Micromegas detector, is shown in figure \ref{fig:DB} (left). This description implies just one event connection and two big charge depositions at the track edges. These two features allow in the case of Micromegas detectors in xenon-quencher mixtures to reduce the background level 4 orders of magnitude in the region of interest, down to 10$^{-4}$~c/keV/kg/yr for a 2 cm-thick copper vessel \cite{Iguaz:2010fi}, as shown in figure \ref{fig:DB} (right).

\subsection{n\_TOF: a neutron flux monitor and 2D profiler}
\begin{figure}[htb!]
\centering
\includegraphics[width=70mm]{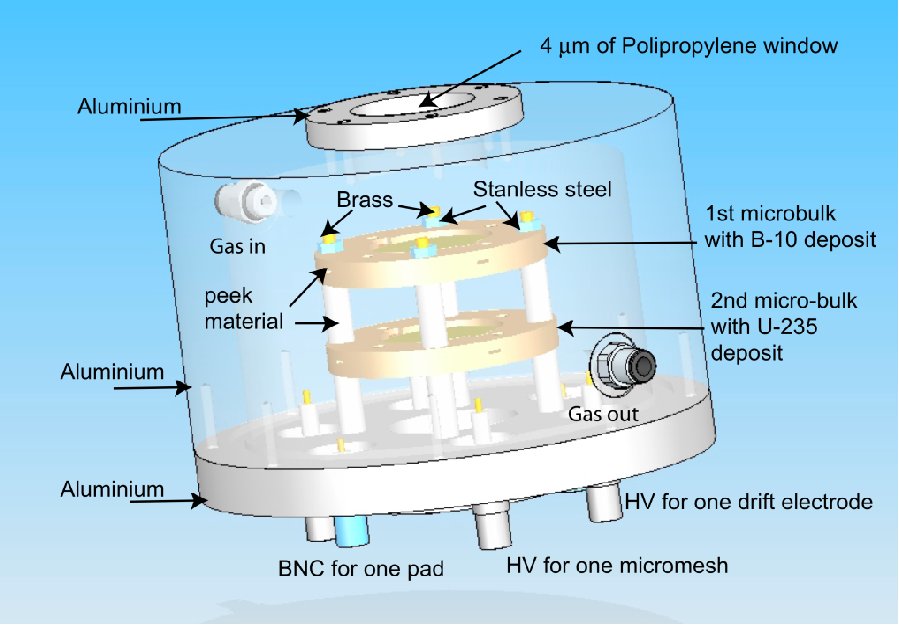}
\hspace{10mm}
\includegraphics[width=70mm]{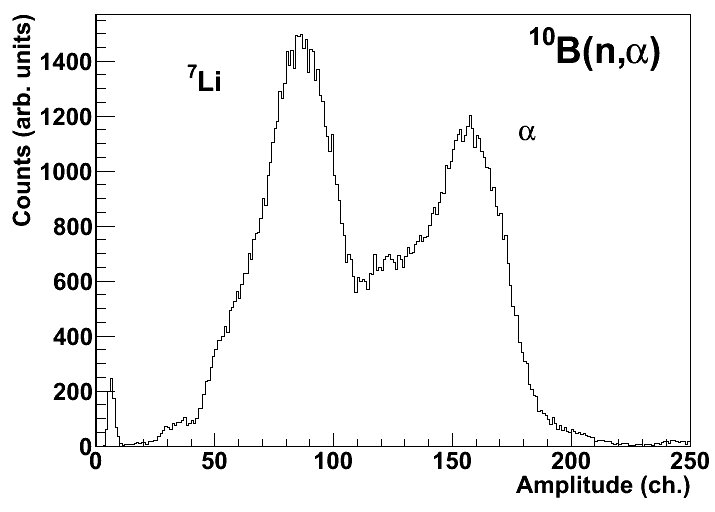}
\caption{\it Left: Schema of the Microbulk based neutron flux monitor. Right: Amplitude spectrum of $^{10}$B generated by the flux monitor.}
\label{fig:nTOF}
\end{figure}

The n\_TOF experiment \cite{Rubbia:1998cr, Gunsing:2007fg} is performing high precision cross-section measurements of neutron-induced reactions, using a neutron beam of a very wide energy range (from thermal up to GeV). Four micromegas detectors have been installed in this experiment, three of them of the microbulk type. The first one is a flux monitor \cite{Adriamonje:2010sa2}, whose schema is shown in figure \ref{fig:nTOF} (left). The system consists of two Micromegas detectors placed inside a cylindrical chamber closed at the ends by kapton foils of 75 $\mu$m thick facing two neutron converters: $^{235}$U (for energies over 1 MeV) and $^{10}$B (for lower energies). As the system is placed at the neutron beam, the detectors have very low mass to minimize the background induced to the other detectors. The detector took data in n\_TOF in 2009 and showed high resistance to radiation, stable gain and good energy resolution. These features allowed it to separate the reaction fragments (figure \ref{fig:nTOF}, right, in the case of a $^{10}$B conversion target). The neutron flux measured was similar to the one obtained during the first phase of the experiment \cite{Colonna:2011nc}. The other Micromegas detectors are used for other goals. The second microbulk detector is used for the measurement of the $^{242}$Pu(n,f) and $^{240}$Pu(n,f) cross sections. The third one is combined with a Total Absorption Calorimeter (TAC) for simultaneous measurement of capture and fission reactions of $^{235}$U. And the last Micromegas detector, which is a bulk type, is a neutron beam profiler, which gives a beam image.

\section{Conclusions}
\label{sec:con}
Microbulk is a Micromegas technology which offers uniform and flexible structures with an excellent energy and time resolution, low background levels and low mass. Two fabrication techniques have been developed (conventional and pillars), where only differ in the quantity of kapton left below the mesh. One detector of each type has been characterized in argon-isobutane mixtures at atmospheric pressure, using a $^{55}$Fe source. Both readouts show similar performances in gain, electron transmission and energy resolution. Microbulk detectors have been also tested in neon-isobutane mixtures to increase the sensitivity in the sub-keV region for applications where the energy threshold is crutial like Dark Matter or synchroton radiation detectors. The maximum gain reached before sparks was $10^5$ and the energy resolution was 10.5\%~FWHM at 5.9 keV. The neon escape peak at 910 eV was observed and the energy theshold was situated at 400 eV. Other gases like helium and quenchers like cyclohexane and ethane are being tested. Microbulk technology has been applied in nuclear (n\_TOF) and astroparticles experiments (CAST, NEXT).



\begin{thebibliography}{00}
\bibitem{Giomataris:1995fq} 
  Y.~Giomataris {\it et al.},
  MICROMEGAS: A high - granularity position - sensitive gaseous detector for high particle - flux environments,
  {\it Nucl.\ Instrum.\ Meth.\  A} {\bf 376} (1996) 29.
\bibitem{Giomataris:2006yg}
  I. Giomataris {\it et al.},
  Micromegas in a bulk,
  {\it Nucl. Instrum. Meth. A} {\bf 560} (2006) 405.
\bibitem{Adriamonje:2010sa}
  S.~Adriamonje {\it et al.},
  Development and performance of Microbulk Micromegas detectors,
  {\it JINST} {\bf 5} (2010) P02001.
\bibitem{Cebrian:2011sc}
  S.~Cebrian {\it et al.},
  Radiopurity of Micromegas readout planes,
  {\it Astropart. Phys.} {\bf 34} (2011) 354.
\bibitem{Giganon:2011ag}
  A.~Giganon,
  private communication.
\bibitem{Iguaz:2010fj}
  F.J.~Iguaz {\it et al.},
  New results of microbulk detectors,
  Talk in the 5th RD51 Collaboration Metting.
\bibitem{Chefdeville:2009mc}
  M. Chefdeville,
  Development of Micromegas-like gaseous detectors using a pixel readout chip as collecting anode,
  PhD Thesis, University of Amsterdam, 2009.
\bibitem{Bronic:1998ikb}
  I.K.~Bronic and B.~Grosswendt,
  Gas amplification and ionization coefficients in isobutane and argon-isobutane mixtures at low gas pressures,
  {\it Nucl. Instrum. Meth. B} {\bf 142} (1998) 219.
\bibitem{Schindler:2010hs}
  H.~Schindler {\it et al.},
  Calculation of gas gain fluctuations in uniform fields,
  {\it Nucl. Instrum. Meth. A} {\bf 624} (2010) 78.
\bibitem{Blum:1994wb}
  W. Blum and L. Rolandi,
  Particle Detection with Drift Chambers,
  Springer-Verlag Berlin Heidelberg, 1993.
\bibitem{Agrawal:1988pca}
  P.C.~Agrawal and B.D.~Ramsey,
  Use of propane as a quench gas in argon-filled proportional counters and comparison with other quench gases,
  {\it Nucl. Instrum. Meth. A} {\bf 273} (1988) 331.
\bibitem{KZioutas:2005kz}
  K.~Zioutas {\it et al.},
  First results of the 2 Axion Solar Telescope (CAST),
  {\it Phys. Rev. Letters} {\bf 94} (2005) 121301.
\bibitem{Andriamonje:2010sa2}
  S.~Andriamonje {\it et al.},
  An improved limit on the axion-photon coupling from the CAST experiment,
  {\it JCAP} {\bf 4} (2007) 010.
\bibitem{EArik09}
  E.~Arik {\it et al.},
  Probing eV-scale axions with CAST,
  {\it JCAP} {\bf 02} (2009) 008.
\bibitem{SAune:2009sa}
  S.~Aune {\it et al.},
  An ultra-low background detector for axion searches,
  {\it J. Phys. Conf. Ser.} {\bf 179} (2009) 012015.
\bibitem{Tomas:2011at}
  A.~Tom\'as {\it et al.},
  CAST micromegas background in the LSC,
  TIPP 2011, Chicago, 2011.
\bibitem{Iguaz:2011fi}
  F.J.~Iguaz {\it et al.},
  The discrimination capabilities of Micromegas detector at low energy,
  TIPP 2011, Chicago, 2011.
\bibitem{Cebrian:2010sc2}
  S.~Cebrian {\it et al.},
  Micromegas readouts for double beta decay searches,
  {\it JCAP} {\b 10} (2010) 1010.
\bibitem{Dafni:2010td}
  T.~Dafni {\it et al.},
  Micromegas planes for the neutrinoless double beta decay search with NEXT,
  {\it J. Phys. Conf. Ser.} {\bf 309} (2011) 012009.
\bibitem{Iguaz:2010fi}
  F.J.~Iguaz,
  Development of a time projection chamber prototype with micromegas technology for the search of the double beta decay of $^{136}$Xe,
  PhD thesis, University of Zaragoza, 2010,
  {\it Preprint:} http://zaguan.unizar.es/record/5731.
\bibitem{Granena:2011fg}
  F.~Gra\~nena {\it et al.},
  NEXT: a HPGXe TPC for neutrinoless double beta decay experiments,
  {\it Preprint:} arXiv:0907.4054.
\bibitem{Balan:2011cb}
  C.~Balan {\it et al.},
  Micromegas operation in high pressure xenon: charge and scintillation readout,
  {\it JINST} {\bf 6} (2011) P02006.
\bibitem{Rubbia:1998cr}
  C.~Rubbia {\it et al.},
  A High Resolution Spallation Driven Facility at the 2-PS to Measure Neutron Cross Sections in the Interval from 1 eV to 250 MeV: a Relative Performance Assessment,
  2-LHC-98-002-EET-Add.1 (1998).
\bibitem{Gunsing:2007fg}
  F.~Gunsing {\it et al.},
  Status and outlook of the neutron time-of-flight facility n\_TOF at 2,
  {\it Nucl. Instrum. Meth. B} {\bf 261} (2007) 925.
\bibitem{Adriamonje:2010sa2}
  S.~Adriamonje {\it et al.},
  A transparent detector for n\_TOF neutron beam monitoring,
  Internal note: n\_TOF-CONF-2010-009.
\bibitem{Colonna:2011nc}
  N.~Colonna {\it et al.},
  Neutron measurements for advanced nuclear systems: The n\_TOF project at 2,
  Accepted in {\it Nucl. Instrum. Meth. B}.
\end{thebibliography}
\end{document}